\documentclass[prb,twocolumn,showpacs,superscriptaddress]{revtex4}
\usepackage{epsfig}
\usepackage{amsmath,amssymb}
\usepackage{graphicx}
\usepackage{footnote}

\usepackage{color}

\usepackage{dsfont}
\usepackage[all]{xy}

\newcommand{\vp}{\varphi}
\newcommand{\vpb}{\bar{\varphi}}
\newcommand{\eb}{\bar{\eta}}
\newcommand{\nub}{\bar{\nu}} 

\newcommand{\Dvp}{\mathcal{D}(\vpb,\vp)}
\newcommand{\Dnu}{\mathcal{D}(\nub,\nu)}

\newcommand{\ceq}[1]{Eq.~(\ref{#1})}
\newcommand{\cfg}[1]{Fig.~\ref{#1}}

\newcommand{\Sd}{\dot{\Sigma}}

\begin{document}

\title{Correlated starting points for the functional renormalization group}
\author{N. Wentzell,$^{1,2,3}$ C. Taranto,$^{3}$ A. Katanin,$^{4,5}$ A. Toschi,$^{3}$ and S. Andergassen$^{1,2}$\\
{\small\em $^1$Faculty of Physics, University of Vienna, Boltzmanngasse 5, 1090 Vienna, Austria} \\
	{\small\em \mbox{$^2$Institut f\"ur Theoretische Physik and CQ Center for Collective Quantum Phenomena}, \\ 
		\mbox{Universit\"at T\"ubingen, Auf der Morgenstelle 14, D-72076 T\"ubingen, Germany}}\\
  {\small\em \mbox{$^3$Institute for Solid State Physics, Vienna University of Technology, 1040 Vienna, Austria}} \\		 
				 {\small\em \mbox{$^4$Institute of Metal Physics, Kovalevskaya Str., 18, 620990, Ekaterinburg, Russia}}\\
				 {\small\em \mbox{$^5$Ural Federal University, 620002, Ekaterinburg, Russia}}
}
\date{\small\today}
\begin{abstract}
We present a general frame to extend functional renormalization group (fRG) based computational schemes by using an exactly solvable {\it interacting} reference problem as starting point for the RG flow. The systematic expansion around this solution accounts for 
a non-perturbative inclusion of correlations. Introducing auxiliary fermionic fields by means of a Hubbard-Stratonovich transformation, 
we derive the flow equations for the auxiliary fields and determine the relation to the conventional weak-coupling truncation of the hierarchy of flow equations. 
As a specific example we consider the 
dynamical mean-field theory (DMFT) solution as reference system, and discuss the relation to the 
recently introduced DMF$^2$RG and the 
dual-fermion formalism. 
\pacs{71.10.-w, 71.27.+a} 
\end{abstract}
\maketitle

\section{Introduction}

One of the main challenges in non-relativistic quantum
many-body theory 
is the
development of powerful tools for treating correlations between
fermionic particles, not limited to specific parameter regimes.
These would provide the keys for understanding and controlling many
of the most exciting experiments currently performed in solid-state, nanoscopic- and cold-atoms
physics. In fact, the state-of-the-art theoretical tools allow for
an accurate treatment of quantum many-body correlations in specific
cases, but their reliability is not guaranteed in general.

A very powerful method, among those currently available and widely used for performing model and realistic calculations of correlated fermions, is arguably the functional renormalization group (fRG) \cite{Salmhofer2007, Berges2002, Kopietz2010, Metzner2012}.  
The starting point of the fRG 
is an exact
functional flow equation, which parametrizes the gradual evolution
from an exactly solvable initial action ${\cal S}_{\rm ini}$ (typically of an uncorrelated problem)
to the full final action ${\cal S}_{\rm fin}$ of the many-body problem of interest.
Expanding the functional flow equation yields an exact infinite
hierarchy of flow equations for the $n$-particle one-particle
irreducible (1PI) vertex functions. However, for most calculations, the hierarchy of equations is truncated at the two-particle level. 
Because of this approximation, the validity of the conventional fRG is  
limited to the perturbative weak-coupling regime, except for situations in which phase space restrictions suppress
higher order contributions\cite{Salmhofer1998, Salmhofer2001,Salmhofer2007}.
For the same reason, the accuracy of the final results depends on the choice of the initial conditions. 

In spite of the limitation to the weak-coupling regime, the fRG has
led to powerful new approximation schemes: In fRG, infrared singularities can be
dealt with much more efficiently than within the traditional resummations
of perturbation theory, due to the built-in RG structure. Moreover, 
--differently from other perturbative approaches, such as RPA--  fRG is
"channel-unbiased": The fRG flow equations include the contributions
of {\sl all} scattering channels (e.g., spin, charge,
particle-particle) and their reciprocal interplay.

The development of novel computation schemes for extending 
the advantages of an fRG-treatment to the strong-coupling regime, where e.g.
Mott-Hubbard metal-insulator transitions can occur,
represents a very challenging, but highly rewarding task.
In fact, the potential of extending the fRG
to the strong-coupling (SC) regime in order to overcome the main restriction of
the conventional implementations has already motivated first pioneering studies\cite{Taranto2014,Kinza2013,Kinza2013a}.
The underlying idea is to access the SC regime by changing the initial conditions of the fRG flow: If these are extracted from the exact solution of a suitably chosen interacting reference problem, a significant part of the correlation effects are included non-perturbatively already from the very beginning, while the remaining ones will be generated, in all scattering channels, by the fRG flow. Formally, this corresponds to taking a SC "reference" system ${\cal S}_{R}$ as an initial action ${\cal S}_{\rm ini}$, provided that it allows for a reliable (numerical or analytical) solution. 
In the case of the Anderson impurity model for example, the atomic-limit and extensions thereof have recently been used as a reference system to define a SC starting point for the fRG flow\cite{Kinza2013, Kinza2013a}. For the Hubbard-model on the other hand, the effective AIM determined self-consistently by the dynamical mean-field theory (DMFT) was chosen to define the initial conditions of the fRG flow\cite{Taranto2014}. This approach, coined DMF$^2$RG, aims at a systematic and channel-unbiased inclusion of correlations,\cite{Taranto2014} beyond the purely local ones described, non-perturbatively, by the DMFT.
We note that the idea of choosing a SC (or non-perturbative) reference system for the fRG flow has been recently introduced also in the context of spin-models\cite{Rancon2014, Reuther2014} or for systems of correlated bosons\cite{Rancon2011, Rancon2011a}.

Irrespective of the performance in specific cases, all extensions of
the conventional fRG face the challenge of proving the validity of the
truncation procedures in the non-perturbative SC regime.
This subject has never been explicitly addressed and calls for a systematic derivation. 

The main goal of this paper is to define the properties of the fRG schemes with a non-perturbative starting point within a rigorous framework and a unified formalism. To this aim, we consider a general starting point for the perturbative expansion: performing a Hubbard-Stratonovich transformation for the fermionic degrees of freedom {\sl not} associated to the chosen (SC) reference system, we derive an action in terms of the auxiliary Hubbard-Stratonovich fermionic fields (referred as "dual" fermions\cite{Rubtsov2008,Brener2008,Hafermann2008,Rubtsov2009,Yang2011,Katanin2013,Kirchner2013,Munoz2013,Antipov2014} in the context of the diagrammatic extensions\cite{Hague2004,Toschi2007,Held2008,Rubtsov2008,Brener2008,Hafermann2008,Rubtsov2009,Valli2010,Yang2011,Katanin2013,Kirchner2013,Munoz2013,Rohringer2013,Antipov2014,Li2014,Valli2014} of DMFT). The resulting equations are compared to those derived by directly working at the level of the physical fermions, and finally, the physical contents underlying the approximations made in the different schemes are critically analyzed.    

The paper is organized as follows. In Sec. II we briefly review the
general procedure for decoupling the fermionic degrees of freedom
associated to a given reference system with the introduction of
auxiliary fermions. The formulation in terms of auxiliary fields and its physical interpretation is presented in Sec. III.  In Sec. VI we derive the flow equations in the auxiliary space and 
in Sec. V we discuss the relation to other methods. Finally, in Sec.\ VI, we summarize our results and draw conclusions.

\section{Interacting reference system}\label{sec:Ref}
In the following, we present a general formalism that allows for an expansion around an interacting reference system. It was first introduced to set up an expansion of the $d$-dimensional Hubbard model around the atomic limit \cite{Pairault1998}, and has recently been used in the dual-fermion (DF) formalism \cite{Rubtsov2008,Brener2008,Hafermann2008,Rubtsov2009,Yang2011,Katanin2013,Kirchner2013,Munoz2013,Antipov2014} to include non-local correlations beyond DMFT.

Let us start with a system of fermionic particles described by a general action of the form
\begin{align}
\mathcal{S}(\vpb,\vp)&= -( \vpb, g^{-1} \vp) + \mathcal{S}_{\rm{int}}(\vpb,\vp).
\label{phys_act}
\end{align}
Here we use the compact notation $(\vpb, \psi) := \sum_{\xi} \vpb_\xi \psi_\xi$. $g$ denotes the non-interacting Green function, and $\mathcal{S}_{\rm{int}}$ contains quartic interaction terms in the Grassmann fields $\vp_\xi, \vpb_\xi$. The multi-index $\xi=(\omega_n,s)$ consists of a fermionic Matsubara frequency $\omega_n$ and a general quantum number $s$ including, e.g., momentum, spin and orbital index. 
We introduce an interacting reference system described by the action
\begin{align}
\mathcal{S}_R(\vpb,\vp)&=-( \vpb, g^{-1}_R \vp) + \mathcal{S}_{\rm{int}}(\vpb,\vp)
\label{ref_act}
\end{align}
written in terms of the same Grassmann fields as action (\ref{phys_act}). It differs from the latter only in the quadratic part $g^{-1}_R$, which is chosen such that the system is exactly solvable.

In order to expand the action
\begin{align}
\mathcal{S}(\vpb,\vp)&=\mathcal{S}_R(\vpb,\vp) - (\vpb, \Delta \vp)
\end{align}
in the difference $\Delta = g^{-1} - g^{-1}_R$ of the quadratic parts, we cannot apply Wick's theorem to the many-particle reference Green functions because the reference action $\mathcal{S}_R$ contains the quartic terms in the fields. Instead, we perform a fermionic Hubbard-Stratonovich transformation
\begin{align}
\hspace{-0.32cm} (\vpb, nD^{-1}n\vp) =  \ln \int \frac{\Dnu}{\operatorname{det}D} e^{-(\nub,D\nu)+(\nub,n\vp)+(\vpb,n\nu)},
\label{hub_strat}
\end{align}
introducing a set of auxiliary fermionic fields $\nu$, $\nub$. We require that the matrices $n_{\xi,\xi'}$ and $D_{\xi,\xi'}$ fulfill the condition $nD^{-1}n=\Delta$, which implies freedom in the choice of $n$.
At this point, we can perform the integration w.r.t.~the physical fields $\vp$, to yield an action 
\begin{equation}
\mathcal{S}_{a}(\nub,\nu)= - (\nub, g^{-1}_{a} \nu) + \hat{\mathcal{V}}(\nub,\nu)
\label{aux_action}
\end{equation}
that depends on the auxiliary fields only (see Appendix B for more details), with an inverse Gaussian propagator
\begin{equation}
g^{-1}_{a} = - n \hspace{0.05cm}\left[ G_R + \Delta^{-1} \right]\hspace{0.05cm}n
\label{f_a_prop}
\end{equation}
that contains correlation effects already through the one-particle Green function of the reference system $G_R$. The interaction of the auxiliary fields reads
\begin{equation}\begin{split}
\hat{\mathcal{V}}(\nub,\nu)=-\left[\ln \mathcal{G}^R(\eb,\eta)+(\eb,G_R\eta)\right]_{\substack{\hspace{-0.19cm}\eta=n\nu\\ \eb=n^{T}\nub}}
\label{int_a}
\end{split}\end{equation}
where $\mathcal{G}_R$ corresponds to the generating functional of the reference system Green functions
\begin{equation}
\mathcal{G}_R(\eb,\eta)=\frac{1}{Z_R}\int \Dvp e^{-[S_R(\vpb,\vp)+(\eb,\vp)+(\vpb,\eta)]}
\end{equation}
with source fields $\eta$, $\eb$. While the freedom in the choice of $n$ can be maintained in all following considerations, we focus on the conventional choice\cite{Rubtsov2008}
\begin{equation}
n=G_R^{-1}
\end{equation}
for the sake of simplicity\footnote{The general expressions containing the freedom $n$ can be found in the Appendix.}. 

\begin{figure}
\includegraphics[width=0.35\textwidth]{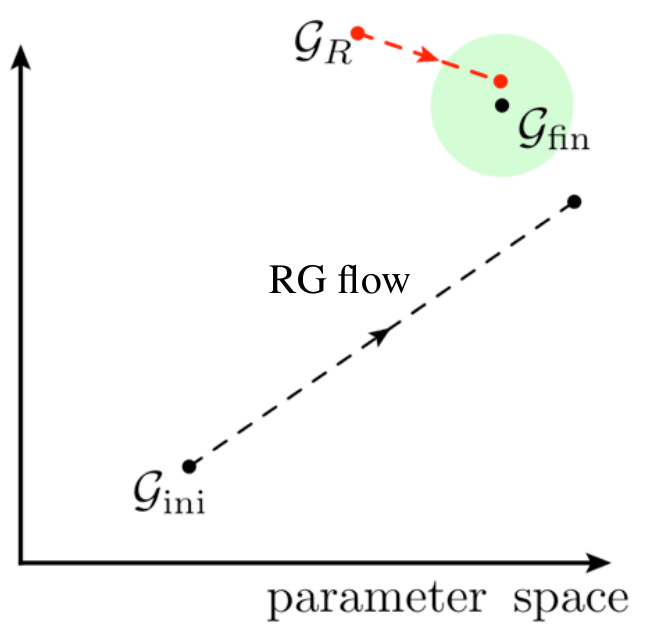}
\caption{Schematic representation of the fRG flow in a general "parameter space". 
In conventional schemes the fRG flow starts from the generating functional of an uncorrelated problem $\mathcal{G}_{\mathrm{ini}}$. The approximations due to the truncation lead to deviations of the result at the end of the flow from the exact final generating functional $\mathcal{G}_{\mathrm{fin}}$. This truncation error can be reduced by using a correlated reference system (provided its generating functional $\mathcal{G}_R$ is exactly known) as a  starting point for the flow.}
\label{sc_schemes}
\end{figure}

Before providing more details about the treatment of the SC problem in auxiliary space, let us briefly sketch in Fig.~\ref{sc_schemes} the idea motivating the formulation of an fRG flow from a SC starting point.
In contrast to the case of the conventional fRG, the initial uncorrelated generating functional $\mathcal{G}_{\mathrm{ini}}$ 
is replaced by the one of the (solvable) interacting system, $\mathcal{G}_R$. A suitable choice reduces the effects of the truncation of the flow equations on the final result, which is therefore closer to the desired $\mathcal{G}_{\mathrm{fin}}$.

\section{The auxiliary problem}\label{sec:aux}
Now that we have reformulated our initial problem in terms of the auxiliary fields, we derive the expansion
around the reference system solution. By definition the reference system solution is contained
completely in the non-interacting theory ($\hat{\mathcal{V}}=0$) of the auxiliary fields, while any treatment of the auxiliary interaction introduces corrections that take into account the solution of the physical system. In the following, we discuss the physical interpretation of the propagator and of the interaction of the auxiliary fields, before introducing the relations between the physical quantities in auxiliary and physical space.

\subsection{Gaussian propagator}\label{sec:prop}
The Gaussian propagator of the auxiliary fields (\ref{f_a_prop})	
has interesting properties. We find that as $\Delta$ tends to zero, $g_a$ vanishes linearly\cite{Hafermann2009}
\begin{equation}
g_a \xrightarrow[\Delta\rightarrow 0]{}G_R\Delta G_R.
\end{equation}
This means that any expansion in the interaction converges asymptotically, as higher order diagrams with $N$ internal lines will be suppressed as $\Delta^{N}$.
For a physical intuition of the propagation described by $g_a$, we write
\begin{equation}\begin{split}
g_a &= \frac{1}{g^{-1}-\Sigma_R} - G_R.
\label{rel_g}
\end{split}\end{equation}
The first term can be interpreted as an approximation to the physical Green function (it actually brings the result of the 0th order expansion in $\hat{\mathcal{V}}$), from which we subtract the full reference propagator. In this sense, $g_a$ corresponds to the difference of the interacting propagators. Therefore it does not exhibit the typical $\sim 1/\omega$ behavior at large frequencies, but rather $\sim 1/\omega^2$.

\subsection{Interaction}\label{sec:int}
By integrating out the physical fields $\vp$, the interaction of the auxiliary fields (\ref{int_a}) is generated. It contains two- and multi-particle interactions that are given by the connected reference Green functions, where $G_R$ is amputated at each external leg.
\begin{widetext}
 We can thus schematically write
\begin{align}\nonumber
\hat{\mathcal{V}}(\bar{\nu},\nu) = &-\frac{1}{4} G^{(2),c}_R \left[\left(G_R^{-1}\nub),\left(G_R^{-1}\nub),(G_R^{-1}\nu\right),(G_R^{-1}\nu\right)\right]
\\ &
 - \frac{1}{9} G^{(3),c}_R  \left[\left(G_R^{-1}\nub),\left(G_R^{-1}\nub),\left(G_R^{-1}\nub),(G_R^{-1}\nu\right),(G_R^{-1}\nu\right),(G_R^{-1}\nu\right)\right] + ...
 \label{new_notation}
\end{align}
where $G^{(m),c}_R$ ($m=2,\ldots,\infty$) denotes the connected $m$-particle reference Green function. In \ceq{new_notation} we introduced the notation $V_n[a_1,...,a_n,a'_1,...,a'_n]$ which is a shorthand for:
\begin{equation}\begin{split}
V_n[a_1,...,a_n,a'_1,...,a'_n]_{\xi_1,...,\xi_n,\xi'_1,...,\xi'_n}=(a_1)_{\xi_1,\psi_1}...  \left(a_n\right)_{\xi_n,\psi_n}V_n(\psi_1,...,\psi_n,\psi_1',...,\psi_n')\left(a'_1 \right)_{\psi_1',\xi_1'}  ... \left(a'_n\right)_{\psi_n',\xi_n'},
\end{split}\end{equation}
where we sum over repeated indices. $V_n$ represents a generic $n$-particle vertex function (e.g., connected Green's function, 1PI vertex ...) and $a_i$ is a two dimensional matrix in the multi index $\xi_i$.

Since the treatment of infinitely many multi-particle interactions poses an impossible task, approximations to the infinite series defining the interaction have to be devised. We drop the interaction terms beyond the quartic one and thus
\begin{eqnarray}\begin{split}\hspace{-0.3cm}
\label{drop}
\hat{\mathcal{V}}(\bar{\nu},\nu) \approx -\frac{1}{4} G^{(2),c}_R \left[\left(G_R^{-1}\nub\right),\left(G_R^{-1}\nub\right),\left(G_R^{-1}\nu\right),\left(G_R^{-1}\nu\right)\right]   &=& -\frac{1}{4} \gamma_2^{R}[\nub,\nub,\nu,\nu],
\end{split}\end{eqnarray}
where $\gamma_2^R$ denotes the one-particle irreducible (1PI) two-particle vertex of the reference system.
\end{widetext}
This represents an approximation based on the fundamental assumption that the effects of 
$(m\geq 3)$-particle scattering processes beyond the description of the reference system can be neglected.
The impact this has on the resulting flow equations is discussed in Sec.~\ref{sec:flow}.
While previous works\cite{Pairault1998,Rubtsov2008,Brener2008,Hafermann2008,Rubtsov2009,Yang2011,Katanin2013,Kirchner2013,Munoz2013,Rohringer2013,Antipov2014} 
have treated the auxiliary interaction $\hat{\mathcal{V}}$ by means of perturbation theory or ladder approaches, we propose using the fRG\cite{Metzner2012} to perform a channel-unbiased resummation of diagrams to all orders in a scale-dependent fashion, thereby further improving on the physical results.

\subsection{Relation to physical quantities}\label{sec:rel}

Once the solution in the auxiliary space is obtained, we need to translate Green- and vertex-functions from the auxiliary to the physical space. These relations can be formulated in a very general way by establishing the connection between the generating functionals. As shown in Appendix C, the generating functionals of the Green functions fulfill
\begin{equation}\begin{split}
\mathcal{G}(\eb,\eta)&=\mathcal{G}_a\left(G_R^{-T}\Delta^{-T}\eb,G_R^{-1}\Delta^{-1}\eta\right) \times e^{-(\eb,\Delta^{-1} \eta)},
\label{genG}
\end{split}\end{equation}
with $G_R^{-T} = \left(G_R^{-1}\right)^{\rm T}$.
Further relations for the generating functional of the connected Green functions, or the effective interaction and the respective derivations are presented in Appendix C. By taking the derivative of \ceq{genG} w.r.t.~the source fields we find the relation between the physical and auxiliary Green functions
\begin{align}
G=\Delta^{-1} + \Delta^{-1} G_R^{-1} G_a G_R^{-1} \Delta^{-1}.
\label{rel_G}
\end{align}
Translated to the self energy, this relation reads
\begin{align}
\Sigma = \Sigma_R + \frac{\Sigma_a}{\mathds{1} + G_R \Sigma_a},
\label{rel_sig}
\end{align}
where the fraction is to be understood as multiplication by the inverse from the right.
The corresponding relation for the 1PI two-particle vertex reads (see Appendix C for further details)
\begin{equation}\begin{split}
\gamma_2 =  \gamma_{2,a}[\zeta,\zeta,\bar \zeta,\bar \zeta],
\label{rel_gam}
\end{split}\end{equation}
with $\zeta=\bar{G}^{-1} G_R$ and $\bar \zeta= G_R\bar{G}^{-1}$, where 
\begin{equation}\bar{G} = \left(g_R^{-1}-\Sigma\right)^{-1}. \label{gbar} \end{equation}
In the following we refer to Eqs.~(\ref{rel_g}), (\ref{rel_sig}) and (\ref{rel_gam}) and generalizations thereof for higher order vertices as the transformation to the auxiliary fields $\mathcal{T}$.

\subsection{DMFT as reference system}\label{sec:dmft}
To  make the procedure described above more concrete, let us focus on the example of a reference system obtained by DMFT. 
This allows for a direct comparison with the DF approach and the recently introduced DMF$^2$RG,\cite{Taranto2014} which is summarized in Sec. \ref{sec:rtom}. 

DMFT can be considered the quantum extension of classical mean field theory \cite{Georges1992,Georges1996} as it can be formally derived as the exact solution of a quantum lattice Hamiltonian in the limit of infinite spatial dimensions ($d\rightarrow \infty$) \cite{Metzner1989}.
In DMFT all  non-local spatial correlations are averaged out, and one can reduce the study of a quantum lattice problem to a self consistently determined local impurity problem. 
For instance, one can consider the action of a one-band Hubbard model \cite{Hubbard1963} given by
\begin{align}
	\mathcal{S}_{\rm lattice} &=T \sum_{\omega_n,\sigma} \int d{\mathbf k}\phantom{1}  \vpb_{\mathbf{k},\sigma}(\omega_n)g^{-1}_{\mathbf{k}}(\omega_n) \vp_{\mathbf{k},\sigma}(\omega_n) \nonumber \\ &\quad+  U \sum_{i}\int d\tau\phantom{1} (n_{i,\uparrow}(\tau) - 1/2) (n_{i,\downarrow}(\tau) - 1/2),
	\label{slattice}
\end{align}
with $T$ being the temperature, $ g_{\mathbf{k}}^{-1}(\omega_n) = (i\omega_n - \mu - \epsilon_{\mathbf{k}})$, $\epsilon_{\mathbf{k}}$ the energy dispersion of the lattice, $i$ the lattice site index, $U$ the Hubbard interaction and $n_{i,\sigma}(\tau)= \vpb_{i,\sigma}(\tau)\vp_{i,\sigma}(0)$.   
 The on-site (local) properties of this action are studied in DMFT by singling out a lattice site and embedding it in an effective bath which accounts for the presence of all the other sites, i.e., an Anderson impurity model (AIM) in an effective bath. To guarantee that the AIM approximates the local physics of the lattice, the effective bath (or hybridization function) $\Gamma(\omega_n)$ has to be computed self-consistently, and the resulting frequency dependence of the effective bath accounts for all purely local quantum correlations. 
The self-consistency condition that determines the effective bath and the propagator $g^{-1}_{\rm{imp}}(\omega_n)=i\omega_n-\Gamma(\omega_n)$ (often referred to as dynamical Weiss field in the DMFT literature) of the AIM reads: 
\begin{equation}
\label{eq:sc}
G_{\rm{DMFT}}=\int d\mathbf{k} \hspace{0.1cm} \left( g_{\mathbf{k}}^{-1} - \Sigma_{\rm{DMFT}}\right)^{-1} = \left(g_{\rm{imp}}^{-1} 
- \Sigma_{\rm{DMFT}}\right)^{-1},
\end{equation}
where $\Sigma_{\rm{DMFT}}$ is the self energy of the self-consistent impurity problem defined by $g_{\rm{imp}}$, and $G_{\rm{DMFT}}$ represents the DMFT approximation to the local interacting lattice Green's function. The self-consistency equation (\ref{eq:sc}) follows directly from the DMFT {\sl assumption} of locality of the lattice self energy, which is clearly an approximation in finite-dimensional systems. 

Since the AIM can be solved exactly, the action of a collection of disconnected self-consistent AIMs, one for each lattice site, is well suited as reference action to approximate the physical action (\ref{phys_act})
\begin{align}	\mathcal{S}_R &= T \sum_{i,\omega_n,\sigma} \vpb_{i,\sigma}(\omega_n)g^{-1}_{\rm{imp}}(\omega_n)\vp_{i,\sigma}(\omega_n) \nonumber \\ &\quad+  U \sum_{i}\int d\tau\phantom{1} (n_{i,\uparrow}(\tau) - 1/2) (n_{i,\downarrow}(\tau) - 1/2).
\label{eq:refaim}
\end{align}
This way, the local physics of the system, computed at the DMFT level, is already included in the reference action. 
Note that assuming the action (\ref{eq:refaim}) as reference action is what is typically\footnote{Let us note that in principle, one can assume the propagator of any AIM to define the action (\ref{eq:refaim}), it does not need to be the self-consistent one of DMFT.} done in the DF\cite{Rubtsov2008,Brener2008,Hafermann2008,Rubtsov2009,Yang2011,Katanin2013,Kirchner2013,Munoz2013,Antipov2014} approaches.
In some cases, also the solution of a cellular DMFT \cite{Kotliar2001} or a Dynamical Cluster Approximation \cite{Hettler2000,Maier2005} calculation has been taken as a reference system \cite{Slezak2009,Yang2011}. This way one is able to include non-perturbatively in the initial conditions also the short-range spatial correlations, providing a complementary multiscale \cite{Slezak2009} framework to treat correlations beyond DMFT over all length scales.   
Hence, the DF reference system is defined by the momentum-independent propagator
$g_{\rm{imp}}$, 
while its interacting Green's function is the local DMFT one
\begin{equation}
G_R^{-1} = g_{\rm{imp}}^{-1} 
- \Sigma_{\rm{DMFT}}.
\end{equation}
Performing the Hubbard-Stratonovich transformation to introduce the auxiliary fermions relative to this reference system, one obtains for the non-interacting propagator
\begin{equation}
g_a= \frac{1}{i\omega_n - \epsilon_{\mathbf{k}} -\Sigma_{\rm{imp}}} - G_R, 
\label{eq:dpro}
\end{equation}
which explicitly depends on the momentum $\mathbf{k}$ through the lattice dispersion $\epsilon_\mathbf{k}$. The first term in Eq. (\ref{eq:dpro}) represents the DMFT approximation to the lattice Green's function under the assumption of a local self energy. $g_a$ is also referred to as the purely non-local propagator, as the local impurity Green's function is subtracted, and therefore it vanishes by summing over the momenta. 
As for the interaction between the auxiliary fields, this is, according to \ceq{drop}, given by the 1PI two-particle vertex of the AIM\cite{Rubtsov2008}.
This input can be calculated \cite{Rohringer2012,Hafermann2014} to high accuracy within the current numerical solvers for the AIM.

\section{Flow equations}\label{sec:flow}

After having reformulated the initial problem by means of the auxiliary action \eqref{aux_action} we now address the issue of solving this problem using the fRG. This procedure is sketched in the right-hand side of \cfg{dfdiag}. Integrating the flow equations in auxiliary space (which are derived in the following) results in an approximated solution for the auxiliary problem that we can eventually translate back to acquire a physical solution. This scheme is then be compared to the one obtained by deriving fRG flow equations directly in the physical space.

\begin{figure}
\vspace{0.5cm}

\includegraphics[width=0.46\textwidth]{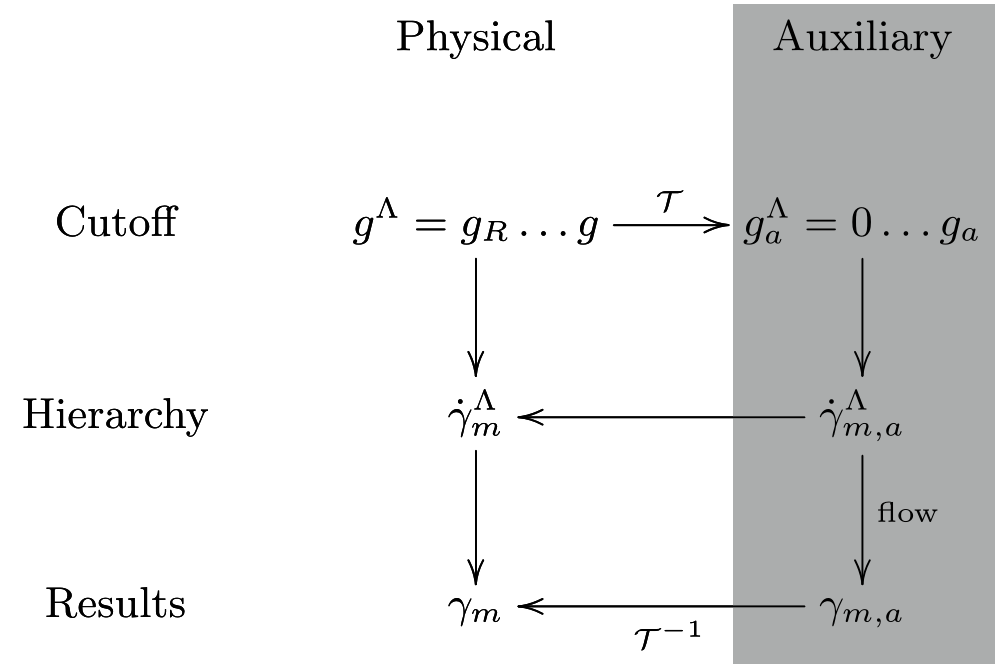}
\caption{Overview of the relation between the flow in physical and auxiliary space; $\mathcal{T}$ denotes the transformation to the auxiliary fields.}
\label{dfdiag}
\end{figure}
\subsection{General formulation}
We recall that the first step to determine the fRG flow equations\cite{Metzner2012} is to substitute the non-interacting propagator $g$ of the system in question by a scale-dependent $g^\Lambda$. This allows for the derivation of an exact functional flow equation that describes the gradual evolution of the effective action as the cutoff scale $\Lambda$ is changed.  Expanding the flow equation in the 1PI vertex functions results in an infinite hierarchy of the form
\begin{equation}
\dot{\gamma}_m^\Lambda = f(\gamma_1^\Lambda, \ldots, \gamma_{m}^\Lambda, \gamma_{m+1}^\Lambda).
\end{equation}
For practical implementations this hierarchy is typically truncated at the two-particle level by assuming that $\gamma_3^\Lambda \approx \gamma_3^{\Lambda_{\rm{ini}}}=0$, reducing the hierarchy to a set of coupled flow equations for the self energy and the 1PI two-particle vertex function
\begin{eqnarray}
\dot{\Sigma}^\Lambda & = & -S^\Lambda \circ \gamma_2^\Lambda
\label{sig_flow}
\\
\dot{\gamma}_2^\Lambda & = & \gamma_2^\Lambda \circ \left( S^\Lambda \circ G^\Lambda + G^\Lambda \circ S^\Lambda \right) \circ \gamma_2^\Lambda.
\label{gam_flow}
\end{eqnarray}
In Eqs.~(\ref{sig_flow}) and (\ref{gam_flow}) the $\circ$ stands for summation over 
internal momenta and quantum numbers according to standard diagrammatic rules as shown explicitly in Appendix A, while 
\begin{equation}
S^\Lambda = -G^\Lambda \left[\partial_\Lambda \left(g^\Lambda\right)^{-1}\right] G^\Lambda=\partial_\Lambda G^\Lambda |_{\Sigma^\Lambda \phantom{1} \mathrm {fix}}
\end{equation}
denotes the single-scale propagator, with 
\begin{equation}
G^\Lambda = \left[ \left(g^\Lambda \right)^{-1} - \Sigma^\Lambda \right]^{-1}.
\end{equation}
The scale dependence of $g^\Lambda$ has to be chosen such that, at the initial scale $\Lambda_{\rm{ini}}$, $\Sigma^{\Lambda_{\rm{ini}}}$ and $\gamma_2^{\Lambda_{\rm{ini}}}$ can be determined exactly, while the physical propagator $g$ is recovered at the final scale $\Lambda_{\rm{final}}$
\begin{equation}
g^{\Lambda_{\rm{final}}} = g.
\end{equation}
In conventional fRG approaches the bare propagator vanishes at the beginning of flow. We introduce the notation
\begin{equation}
g^{\Lambda}= 0 \ldots g
\end{equation}
to describe this 
scale-dependence.
While this choice results in trivial initial conditions that can be read off from the microscopic model action directly, recent works\cite{Kinza2013a,Kinza2013,Reuther2014,Taranto2014} introduced the idea of starting the fRG flow from a reference system solution by choosing 
\begin{equation}
g^{\Lambda}= g_R \ldots g.
\label{str_cpl_frg}
\end{equation}
This approach corresponds to the left-hand side of \cfg{dfdiag}.
\subsection{Flow in the auxiliary space}
In contrast to the derivation in the previous subsection, we now consider the reformulated problem for the auxiliary fields, for which we set up an fRG flow in the conventional sense. For this, we introduce a scale dependent auxiliary field propagator $g_a^\Lambda = 0\ldots g_a$. 
The resulting flow equations then read as Eqs.~(\ref{sig_flow}) and (\ref{gam_flow}), where all physical objects have to be replaced by their auxiliary equivalents. Note that the relations (\ref{rel_sig}) and (\ref{rel_gam}) between the scale-dependent physical and auxiliary quantities remain valid. 
Even though the auxiliary self energy $\Sigma_a^{\Lambda}$ vanishes at the scale $\Lambda_{\rm{ini}}$, the initial conditions are in general highly nontrivial, since $\gamma_{2,a}^{\Lambda_{\rm{ini}}}=\gamma_2^R$.

As mentioned in Sec.~\ref{sec:int}, we approximate the bare auxiliary interaction by its quartic term. This corresponds to neglecting the effect of multi-particle scattering processes when calculating corrections of the reference system solution towards the physical one. In the auxiliary fRG flow, this results in an initially vanishing scale-dependent three-particle vertex, $\gamma_{3,a}^{\Lambda_{\rm ini}}=\gamma^R_{3,c}=0$, which justifies our truncation of the auxiliary flow equation hierarchy.
Note, however, that multi-particle scattering processes are of course included in the exact solution of our reference system, and thus in the initial conditions ($\gamma^{\Lambda_{\mathrm{ini}}}_{2,a}=\gamma^R_{2}$) of the flow. 
After solving these flow equations numerically, results have to be translated back, using the transformation $\mathcal{T}^{-1}$ as described in Sec. \ref{sec:rel}.

This approach is similar in spirit to recent approaches\cite{Kinza2013a,Kinza2013,Reuther2014,Taranto2014} following \ceq{str_cpl_frg}, as depicted in the left-hand side of \cfg{dfdiag}. In fact, any cutoff of the form \eqref{str_cpl_frg} is translated to $g_a^\Lambda=0\ldots g_a$ in auxiliary space by means of 
(\ref{rel_g}).  
Without any approximations the two flow schemes yield identical results. The approximations due to the truncation the flow equation hierarchy however induce important differences, as illustrated in the following. We compare the two paths in \cfg{dfdiag} leading to a hierarchy of physical flow equations $\dot{\gamma}_m^{\Lambda}$. In particular, we will compare the equations for the one- and two-particle 1PI vertex functions as relevant to common truncation schemes. Assuming that the scale dependence $g_a^\Lambda$ translates into $g^\Lambda$ by 
\eqref{rel_g}, we determine the relation between the scale-dependent Green functions. Introducing the scale-dependence in \ceq{rel_G} and solving for $G_a^\Lambda$ yields
\begin{align}
G^\Lambda_a = \zeta^\Lambda \left(G^\Lambda - \bar{G}^\Lambda \right) \bar{\zeta}^\Lambda , 
\end{align}
where $\bar{G}$, defined by \ceq{gbar}, acquires a scale dependence via $\Sigma^\Lambda$. For the single-scale propagator we use the definition $S_a^\Lambda = \partial_\Lambda G_a^\Lambda |_{\Sigma_a^\Lambda \phantom{1} \mathrm {fix}}$ to obtain
\begin{equation}
S_a^\Lambda = \zeta^\Lambda S^\Lambda \bar{\zeta}^\Lambda.
\end{equation}
Considering that 
\begin{equation}
\gamma_{2,a}^\Lambda =  \gamma_{2}^\Lambda\left[\left( \zeta^\Lambda \right)^{-1},\left( \zeta^\Lambda \right)^{-1},\left(\bar{\zeta}^\Lambda \right)^{-1},\left(\bar{\zeta}^\Lambda \right)^{-1}\right],
\label{tplr}
\end{equation}
we find that each diagram contributing to the auxiliary flow can be translated to its physical counterpart by making the substitutions
\begin{equation}\begin{split}
\gamma_{2,a}^\Lambda &\rightarrow \gamma_2^\Lambda \\
S^\Lambda_a &\rightarrow S^\Lambda \label{trans_d}\\
G^\Lambda_a &\rightarrow G^\Lambda - \bar{G}^\Lambda.
\end{split}\end{equation}
To relate the flow equations for the self energy in the physical and auxiliary space we take the $\Lambda$-derivative of the scale-dependent  \ceq{rel_sig} 
\begin{align} 
\Sd^\Lambda=\zeta^\Lambda \Sd_a^\Lambda \bar{\zeta}^\Lambda.
\label{sig_sig}
\end{align}
Applying the translation rules above, the corresponding flow equation in physical space remains unchanged, and is thus given by \ceq{sig_flow}.
The flow equations for the two-particle vertex, instead, can be obtained by taking the $\Lambda$-derivative of \ceq{tplr}. Besides the contribution arising by a direct translation of the diagrams in auxiliary space through \ceq{trans_d}, additional terms arise due to the derivative of the $\zeta^\Lambda$ factors attached at each leg
\begin{align}
\dot{\gamma}_2^\Lambda &=\dot{\gamma}_{2,a}^\Lambda\left[\zeta^\Lambda,\zeta^\Lambda,\bar\zeta^\Lambda,\bar\zeta^\Lambda\right] - \gamma_{2,a}^\Lambda\left[\dot{\Sigma}^\Lambda G_R,\zeta^\Lambda,\bar\zeta^\Lambda,\bar\zeta^\Lambda\right]\nonumber\\&\quad
- \gamma_{2,a}^\Lambda \left[\zeta^\Lambda,\dot{\Sigma}^\Lambda G_R ,\bar\zeta^\Lambda,\bar\zeta^\Lambda\right]
- \gamma_{2,a}^\Lambda \left[\zeta^\Lambda,\zeta^\Lambda,G_R\dot\Sigma^\Lambda,\bar\zeta^\Lambda\right]
\nonumber\\&\quad- \gamma_{2,a}^\Lambda \left[\zeta^\Lambda,\zeta^\Lambda,\bar\zeta^\Lambda,G_R\dot{\Sigma}^\Lambda\right].
\end{align}
Finally, by truncating the auxiliary flow equation at the one-loop level, we get the following flow equations in physical space

\begin{figure}[b!]
\includegraphics[width=\columnwidth]{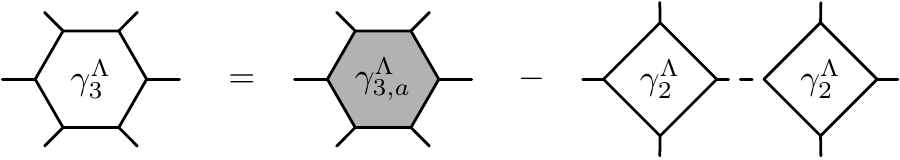}
\caption{Relation between the physical and auxiliary scale-dependent three-particle vertex functions.}
\label{vsplit}
\end{figure}

\begin{widetext}
\begin{align}
\dot{\gamma}_2^\Lambda &=
\gamma_2^\Lambda \circ \left( S^\Lambda \circ G^\Lambda + G^\Lambda \circ S^\Lambda \right) \circ \gamma_2^\Lambda  - \gamma_2^\Lambda \circ \left( S^\Lambda \circ \bar{G}^\Lambda + \bar{G}^\Lambda \circ S^\Lambda \right) \circ \gamma_2^\Lambda \nonumber\\&\quad-\gamma_2^\Lambda\left[\dot{\Sigma}^\Lambda \bar{G}^\Lambda,\mathds{1},\mathds{1},\mathds{1}\right]-\gamma_2^\Lambda\left[\mathds{1},\dot{\Sigma}^\Lambda \bar{G}^\Lambda,\mathds{1},\mathds{1}\right] -\gamma_2^\Lambda\left[\mathds{1},\mathds{1}, \bar{G}^\Lambda\dot{\Sigma}^\Lambda,\mathds{1}\right] \nonumber\\&\quad-\gamma_2^\Lambda\left[\mathds{1},\mathds{1},\mathds{1},\bar{G}^\Lambda\dot{\Sigma}^\Lambda\right] ,
\label{new_gam_flow}
\end{align}
 
\begin{figure}
\includegraphics[width=0.23\textwidth]{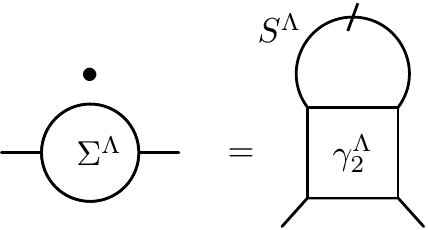}
\vspace{0.4cm}
\includegraphics[width=0.77\textwidth]{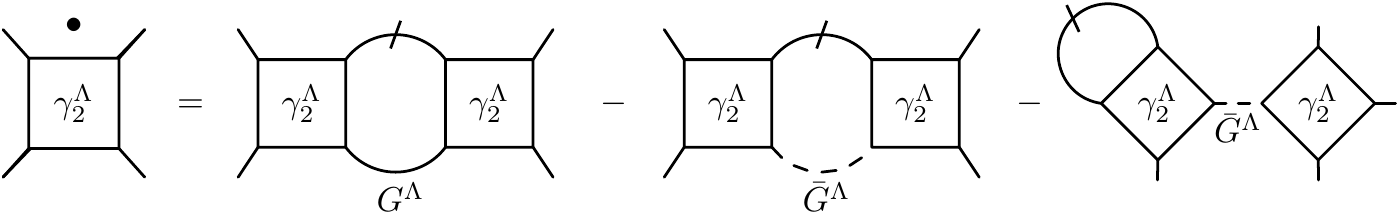}
\caption{Diagrammatic representation of the physical flow equations for $\Sigma^\Lambda$ and $\gamma_2^\Lambda$ that correspond to solving the conventional flow equations for the auxiliary fields using the 2nd order truncation.}
\label{flow_diag}
\end{figure}

\end{widetext}
for a detailed derivation we refer to Appendix D.
The corresponding diagrams are shown in \cfg{flow_diag}. 
The first term of \ceq{new_gam_flow} is identical to the flow equation \ceq{gam_flow} for the two-particle vertex in physical space, while the other terms are not present in the conventional scheme. In particular, we note that the last four terms of \ceq{new_gam_flow} are reducible in $\bar{G}^\Lambda$.
However, this "reducibility" does not necessarily coincide with the reducibility in $G^\Lambda$ and could be attributed to the specific approximation performed here. This can also be understood by considering the relation between the physical and auxiliary scale-dependent three-particle vertices, depicted in \cfg{vsplit}. Assuming that the effect of $\gamma_{3,a}^\Lambda$ on the flow is negligible, we find that the effect of the physical 1PI three-particle vertex can be described 
by the effect of the rightmost term in Fig.~\ref{vsplit}. By connecting two of the six external (amputated) legs of this diagram with a single-scale propagator, one obtains the one-loop correction [second term in \ceq{new_gam_flow}] as well as the "reducible" corrections [last four terms in \ceq{new_gam_flow}] to the conventional flow equation.  
In particular, 1PI three-particle vertex corrections are included under the assumption that the auxiliary 1PI three-particle vertex vanishes. In conventional fRG, where it is assumed that both sides of the equation depicted diagrammatically in \cfg{vsplit} do not contribute, three-particle vertex effects are fully neglected, unless they are explicitly accounted for, e.g., by two-loop diagrams\cite{Katanin2004,Eberlein2014}.
Let us note that a similar diagrammatic structure of the flow equations has been determined in a recent two-band fRG approach\cite{Maier2012}, where a high-energy band was included
perturbatively, resulting in an effective one-particle reducible three-particle interaction of the low-energy band. The corresponding contribution to the two-particle vertex flow was subsequently considered explicitly.
We emphasize that in the present approach the one-particle "reducible" (in the sense explained above) corrections to the physical self-energy inferred by (\ref{rel_sig}) appear at the two-particle level\cite{Katanin2013} (as the last term in the flow equation for the vertex in \cfg{flow_diag}). On the other hand, comparing the auxiliary flow equations to the ones obtained in the recently introduced 1PI approach\cite{Rohringer2013}, allows to 
attribute the one-particle "irreducible" correction of the two-particle vertex to the the first two terms. Translated to the self-energy, for the Hubbard model with the DMFT as a starting point, these terms produce the analogue of the purely non-local contribution to the self-energy, which contains the Green function difference $G-\bar{G}$ in the diagrammatic series. 
In contrast, the contribution containing "irreducible" diagrams with at least one internal $\bar{G}$-line
of the 1PI approach is absent here, since the auxiliary three-particle vertex is neglected. 
For the half-filled $2d$ Hubbard model with the DMFT as a starting point, this approximation is justified 
at relatively strong coupling, where the respective contribution was found to be largely compensated by the contribution of the other channels~\cite{Rohringer2013,Katanin2009}.

\section{Relations to other methods}\label{sec:rtom}
The key idea presented in the previous sections is to approach the physical problem of treating strongly interacting fermions in a channel unbiased way in two essential steps: $i$) Setting up an expansion around a reference system solution by means of auxiliary fermions, and 
$ii$) solving the auxiliary problem by the fRG.
Step $i$), first introduced in Ref. \onlinecite{Pairault1998}, is the basis of all the studies performed in the DF formalism\cite{Rubtsov2008,Brener2008,Hafermann2008,Rubtsov2009,Yang2011,Katanin2013,Kirchner2013,Munoz2013,Antipov2014} as well as of the 1PI approach\cite{Rohringer2013} and  in the study of impurity systems within superperturbation theory\cite{Hafermann2009}. An fRG flow from an interacting starting point, that is step $ii$), has been recently proposed\cite{Kinza2013a,Kinza2013,Rancon2014,Reuther2014,Taranto2014}.

\begin{figure}[b!]
\begin{tabular}{c|c}
Physical flow & Auxiliary flow\\
\hline  \\
$g^\Lambda = g_R \ldots g$ & $g^\Lambda_a = 0 \ldots g_a$ \\
\includegraphics[width=0.1\textwidth]{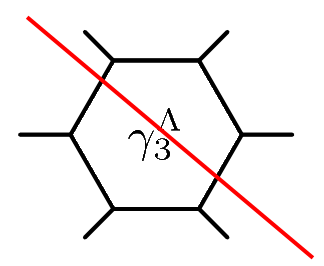} & \includegraphics[width=0.1\textwidth]{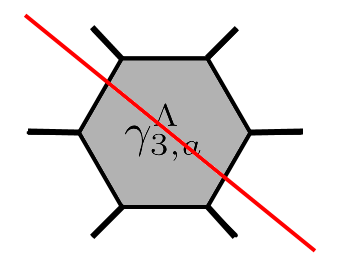} \\
Effects of 1PI three-particle & Mimic 1PI three-particle \\
vertex fully neglected  & vertex by means of\\
& \includegraphics[width=0.25\textwidth]{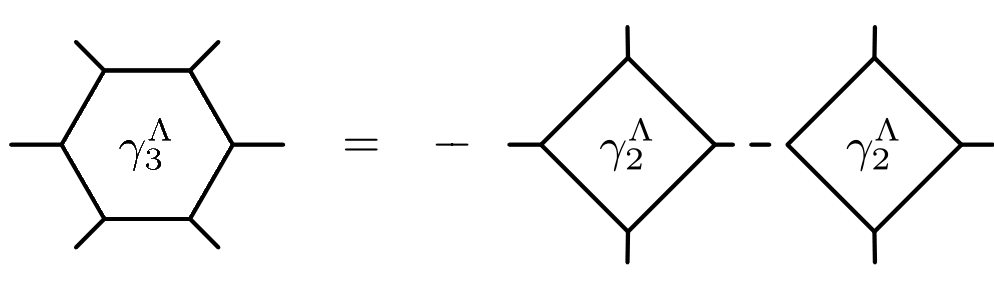}
\end{tabular}
\caption{Effects of the truncation in the physical and auxiliary flow.}
\label{summary_table}
\end{figure}

To be more concrete, in the following, we concentrate on the cases where the DMFT solution is used as a reference system solution, which corresponds to the DF approaches regarding step $i$) and which has been also taken as direct input for the DMF$^2$RG flow. 
This corresponds to take an action of the form of \ceq{eq:refaim} as initial action and constructs the flow to the final action by interpolating the bare propagator from the AIM one, $g_{\rm imp}$, to the final (lattice) one, $g$.
Differently to the auxiliary field method proposed here, however this is done directly in physical space.
The important question to be addressed, then, is whether considering the flow in the auxiliary or in the physical space is more convenient.
In DMF$^2$RG, one is assuming that the effect of the DMFT three-particle 1PI vertex in correcting the self energy of the reference system is small compared to the effect of two-particle one.
In the case of conventional fRG this can be justified, at least at weak coupling, by a power counting argument\cite{Salmhofer2001} showing that the leading order contribution to the three-particle vertex is one order higher in the interaction compared to the leading one in the two-particle vertex.
At intermediate-to-strong coupling, however, this argument does not apply anymore. Hence, in DMF$^2$RG one should rely on the fact that the main contribution of the higher order vertex functions to the truncated fRG flow is already included in the initial condition as sketched in \cfg{sc_schemes}.
Obviously, there is no guarantee that the effect of the three-particle 1PI vertex can be neglected in general.
The difference with the approach discussed here in terms of the auxiliary fields is the following: 
In the present approach, the {\sl auxiliary} three-particle vertex is neglected as discussed in Sec. \ref{sec:flow} (see schematic representation in Fig.~\ref{summary_table}).
At the initial scale this corresponds to neglecting the {\sl connected} reference three-particle vertex\footnote{By connected vertex functions we refer to connected Green functions amputated by the {\sl interacting} one-particle Green function.} instead of the corresponding 1PI one.
From the discussion above, it is clear that the crucial question is whether the physical 1PI three-particle vertex or the auxiliary one has a stronger effect on the corresponding flow. 
This certainly strongly depends on the problem under consideration and requires further focused investigations. Away from weak coupling this question is all but trivial, apart for some special cases, e.g., the Falicov-Kimball model\cite{Antipov2014}. Hitherto, due to its intrinsic numerical complexity, barely any knowledge about the three-particle quantities is available in the literature\cite{Hafermann2010}. A noticeable exception is the case of the Falicov-Kimball model, where it has been shown that the local auxiliary three-particle vertex $\gamma_{3,a}$ exactly vanishes in the particle-hole symmetric case. Here, however, perturbation theory considerations and the application of the Fury theorem \cite{MandlShaw2002} would suggest a simultaneous vanishing of the 1PI three-particle vertex for the physical fields. These results, however, cannot be directly extended to the Hubbard model.
Therefore one has to critically analyze the results obtained with each approach and for each specific case.
In order to compare with DF calculations, one can directly analyze the diagrammatic contributions in auxiliary space where the DF approach\cite{Rubtsov2008,Brener2008,Hafermann2008,Rubtsov2009,Yang2011,Katanin2013,Kirchner2013,Munoz2013,Antipov2014} usually exploits perturbation theory or ladder resummations, as additional approximations. While one can not expect to capture diverging fluctuations by means of simple perturbation theory, ladder approaches are per definition biased towards a selected channel. On the other hand, by treating the auxiliary problem by means of the fRG one is able to treat competing scattering channels in an unbiased way. In particular, our approach allows for an improved computation of the solution of the auxiliary problem, including, although approximately, the parquet-approximation diagrams.
Let us note that, differently from other non-perturbative
schemes\cite{Toschi2007,Held2008,Slezak2009,Valli2010,Valli2014}, the calculations are based on the 1PI two-particle vertex, and do not require the two-particle irreducible (2PI) vertex at any point of the algorithm. 
This way one can circumvent the technical problems arising from the recently shown \cite{Yang2011,Schafer2013,Janis2014} divergencies of the 2PI vertex at low frequencies, which are not associated with any thermodynamic transition.

\section{Conclusion}\label{concl}

We have demonstrated how the theory of an fRG-based expansion around a SC reference system can be rigorously formulated in the general framework of auxiliary fermionic variables. In particular, we have derived the explicit expressions for the fRG flow equations starting from a generic (exactly solvable) SC reference problem in the auxiliary fermionic fields and the corresponding transformation relations to calculate the physical quantities of the final solution. 
These derivations allow to clarify the relation to the
fRG flow equations formulated directly for the physical fermionic fields, including the first pioneering ones reported in the recent literature\cite{Taranto2014,Kinza2013,Rancon2011,Rancon2014,Reuther2014}.
Furthermore, we could also elucidate the implications 
of the approximations introduced by truncating the hierarchy of the flow equations in the different schemes, 
see \cfg{summary_table} for a summary.
This represents indeed a pivotal aspect for all strong-coupling fRG algorithms, since the conventional arguments justifying the truncation
does not hold any longer beyond the weak-coupling regime. Hence, a precise definition of the diagrammatic content associated to the truncation of a strong-coupling fRG flow is essential for adapting the novel algorithms to the non-perturbative physics of interest.   
The reported analytic and diagrammatic results, together with the physical insights which can be captured within the different formulations of the fRG with non-perturbative starting points, will provide an important reference for any future method development in this promising direction.
\section*{Acknowledgments}
We thank T. Costi, A. Eberlein, K. Held, C. Honerkamp, M. Kinza, V. Meden, W. Metzner, G. Rohringer and M. Salmhofer for discussions, and S. Sproch for critically reading the manuscript.
We acknowledge financial support from FWF SFB ViCoM F4104, from DFG through ZUK 63, SFB/TRR21, and FOR 723, and from the Dynasty Foundation, Russia.

\begin{widetext}
\section*{Appendix}
\subsection{Notation details}
\label{ssec:notation_det} 
Here we present explicitly \ceq{sig_flow} and \ceq{gam_flow}, shown in compact notation in the main text.
The flow equation for the self energy reads 
\begin{eqnarray}
\label{explicit}
\dot{\Sigma}(\xi_1;\xi'_1)=-\frac{1}{\beta}\sum_{\omega_2}\sum_{s'_2,s_2} S^\Lambda_{s'_2;s_2}(\omega_2) \gamma_2^\Lambda(\xi_1,\xi_2;\xi'_1,\xi'_2).   
\end{eqnarray}   
We use the convention that for an $m$-particle quantity the first $m$ arguments refer to the outgoing lines, and the last $m$ to the incoming lines. 
The loop-variables are $\xi_2=(\omega_2,s_2)$ and $\xi'_2(\omega'_2,s'_2)$. 
It is implicitly assumed that the energy is conserved ($\omega_2=\omega'_2$), while the multi-index $s_2$ consists of a set of continuous and discrete quantum numbers. Therefore the summations $\sum_{s_2,s'_2}$ have to be understood as integrations or summations respectively. 
For example, in the case where only spin and momentum are considered for a translationally invariant system ($\mathbf{k}_2=\mathbf{k}'_2$), \ceq{explicit} reads 
\begin{eqnarray}
\dot{\Sigma}(\xi_1;\xi'_1)=-\frac{1}{\beta}\sum_{\omega_2}\sum_{\sigma_2,\sigma'_2}\int d\mathbf{k}\phantom{1}S^\Lambda_{\mathbf{k}'_2\sigma'_2,\mathbf{k}_2\sigma_2}(\omega_2)\gamma_2^\Lambda(\omega_1\mathbf{k}_1\sigma_1,\omega_2\mathbf{k}_2\sigma_2;\omega'_1 \mathbf{k}'_1\sigma'_1,\omega'_2\mathbf{k}'_2\sigma'_2).
\end{eqnarray} 
For the 1PI two-particle vertex the flow equation is usually subdivided in three channels (particle-particle, particle-hole direct and particle-hole crossed) corresponding to the diagrammatic contributions 
\begin{equation} 
\dot{\gamma}_2^\Lambda(\xi_1,\xi_2;\xi'_1,\xi'_2)=\mathcal{C}_{pp}(\xi_1,\xi_2;\xi'_1,\xi'_2)+\mathcal{C}_{pp-d}(\xi_1,\xi_2;\xi'_1,\xi'_2)+\mathcal{C}_{ph-c}(\xi_1,\xi_2;\xi'_1,\xi'_2),
\end{equation}  
with 
\begin{eqnarray}
\nonumber  \mathcal{C}_{pp}(\xi_1,\xi_2;\xi'_1,\xi'_2)&=&\frac{1}{\beta}\sum_{\omega_3}\sum_{s_3,s'_3,s _4,s'_4}\Big[S^\Lambda_{s'_3s _3}(\omega_3)G^\Lambda_{s'_4,s _4}(\omega_1+\omega_2-\omega_3)+ \\ &&G^\Lambda_{s'_3s _3}(\omega_3)S_{s'_4,s _4}^\Lambda(\omega_1+\omega_2-\omega_3)\Big] 
 \gamma_2^\Lambda(\xi_1,\xi_2;\xi'_3,\xi'_4)\gamma_2^\Lambda(\xi_3,\xi_4;\xi_1',\xi_2'), \\
\nonumber 
\mathcal{C}_{ph-d}(\xi_1,\xi_2;\xi'_1,\xi'_2)&=&-\frac{1}{\beta}\sum_{\omega_3}\sum_{s _3,s _3',s _4,s _4'}\Big[S^\Lambda_{s'_3s _3}(\omega_3)G_{s'_4,s _4}^\Lambda(\omega_1-\omega'_1+\omega_3)+ \\ &&G^\Lambda_{s'_3s _3}(\omega_3)S_{s'_4,s _4}^\Lambda(\omega_1-\omega'_1+\omega_3)\Big] 
 \gamma_2^\Lambda(\xi_1,\xi_3;\xi'_1,\xi'_4)\gamma_2^\Lambda(\xi_4,\xi_2;\xi'_3,\xi'_2), \\
 \nonumber 
\mathcal{C}_{ph-c}(\xi_1,\xi_2;\xi'_1,\xi'_2)&=&\frac{1}{\beta}\sum_{\omega_3}\sum_{s _3,s _3',s _4,s _4'}\Big[S^\Lambda_{s'_3s _3}(\omega_3)G_{s'_4,s _4}^\Lambda(\omega_1-\omega'_2+\omega_3)+ \\ &&G^\Lambda_{s'_3s _3}(\omega_3)S_{s'_4,s _4}^\Lambda(\omega_1-\omega'_2+\omega_3)\Big] 
 \gamma_2^\Lambda(\xi_1,\xi_3;\xi'_2,\xi'_4)\gamma_2^\Lambda(\xi_2,\xi_4;\xi'_1,\xi'_3). 
\end{eqnarray} 

\subsection{Auxiliary fields}
\label{ssec:aux_fields}
Performing the Hubbard-Stratonovich transformation (\ref{hub_strat}) yields an action
\begin{align}
\mathcal{S}(\vpb,\vp,\nub,\nu)=&\mathcal{S}_R(\vpb,\vp)+ \mathcal{S}_{\vp}(\vpb,\vp,\nub,\nu) + (\nub,n\Delta^{-1}n \nu),
\end{align}
with $\mathcal{S}_{\vp}(\vpb,\vp,\nub,\nu) = (\nub,n\vp) + (\vpb,n\nu)$. We determine the interaction $\hat{\mathcal{V}}$ of the auxiliary $\nu$-fermions by integrating out the $\vp$-fields and obtain
\begin{equation}
\int \Dvp e^{ -\mathcal{S}_R(\vpb,\vp) - \mathcal{S}_{\rm{\vp}}(\vpb,\vp,\nub,\nu)  }
=Z_R~e^{\,(\nub,\tilde{Q}\nu) - \hat{\mathcal{V}}(\nub,\nu)}.
\label{dual_int}
\end{equation}
This relation defines $\hat{\mathcal{V}}$ and $\tilde{Q}$, where $\tilde{Q}$ is to be chosen such that $\hat{\mathcal{V}}$ does not contain any quadratic part in the auxiliary fermions. Note that the l.h.s. is closely related to the generating functional of the reference Green functions
\begin{equation}
\int \Dvp e^{-\mathcal{S}_R(\vpb,\vp) - \mathcal{S}_{\rm{\vp}}(\vpb,\vp,\nub,\nu) } =\left. Z_R \hspace{0.15cm} \mathcal{G}^R(\eb,\eta)\right|_{\eta=n\nu, \eb=n^{T}\nub}.
\end{equation}
Inserting in \ceq{dual_int} and solving for $\hat{\mathcal{V}}$ we obtain
\begin{equation}\begin{split}
\hat{\mathcal{V}}(\nub,\nu)= -\left[\ln \mathcal{G}^R(\eb,\eta)-(\eb,n^{-1}\tilde{Q}n^{-1}\eta)\right]_{\eta=n\nu, \eb=n^{T}\nub} = -\left[-(\eb,G_R\eta) + \mathcal{O}(\eb^2\eta^2)-(\eb,n^{-1}\tilde{Q}n^{-1}\eta)\right]_{\eta=n\nu, \eb=n^{T}\nub}.
\end{split}\end{equation}
For the quadratic part of $\hat{\mathcal{V}}$ to vanish we have to choose $n^{-1}\tilde{Q}n^{-1} = -G_R $ and hence $\tilde{Q} = -n G_R n$. Thus
\begin{equation}\begin{split}
\hat{\mathcal{V}}(\nub,\nu)=-\left[\ln \mathcal{G}^R(\eb,\eta)+(\eb,G_R\eta)\right]_{\eta=n\nu, \eb=n^{T}\nub},
\end{split}\end{equation}
in accordance with \ceq{int_a}.
This functional generates two-particle and multi-particle connected Green functions, where $n$ is appended at the outer legs. The free propagation of the auxiliary fields is then described by
\begin{equation}
Q_a = \tilde{Q} - n\Delta^{-1}n = -n\left[ G_R + \Delta^{-1} \right]n
\end{equation}
as shown in \ceq{f_a_prop} for $Q_a = g_a^{-1}$.

\subsection{Relation between physical and auxiliary space} \label{ssec:rel}
We here relate the physical to the auxiliary Green functions. For this we determine the relation between the generating functional of the physical Green functions
\begin{equation}\begin{split}
\mathcal{G}(\eb,\eta)&= \frac{1}{Z \hspace{0.05cm} \det D} \int \Dvp \hspace{0.1cm} \Dnu \hspace{0.1cm} e^{-\mathcal{S}(\vpb,\vp,\nub,\nu) + (\eb,\vp)+(\vpb,\eta)},
\end{split}\end{equation}
and of the auxiliary Green functions
\begin{equation}\begin{split}
\mathcal{G}_a(\eb,\eta)&= \frac{1}{Z_a Z_R} \int \Dnu  \hspace{0.1cm} \Dvp \hspace{0.1cm} e^{-\mathcal{S}(\vpb,\vp,\nub,\nu) + (\eb,\nu)+(\nub,\eta)}.
\end{split}\end{equation}
We evaluate
\begin{align}
-\mathcal{S}(\vpb,\vp,\nub,\nu) + (\eb,\vp)+(\vpb,\eta) 
= -\mathcal{S}_R(\vpb,\vp) + (\eb + n^{T}\nub,\vp) + (\vpb,\eta+n\nu) - (\nub,n\Delta^{-1}n\nu)
\end{align}
and substitute $n\nu\rightarrow n\nu' - \eta$ and $n^{T}\nub\rightarrow n^{T}\nub' - \eb$
\begin{align}\nonumber
&-\mathcal{S}_R(\vpb,\vp) + (\nub',n\vp) + (\vpb,n\nu') - (\nub',n\Delta^{-1}n\nu') - (\eb,\Delta^{-1} \eta)+ (\eb,\Delta^{-1}n\nu') + (\nub',n\Delta^{-1}\eta)\\
=&-\mathcal{S}(\vpb,\vp,\nub',\nu') - (\eb,\Delta^{-1}\eta) + (n^{T}\Delta^{-T}\eb,\nu') + (\nub',n\Delta^{-1}\eta).
\end{align}
Thus
\begin{equation}\begin{split}
\mathcal{G}(\eb,\eta)&=\mathcal{G}_a\left(n^{T}\Delta^{-T}\eb,n\Delta^{-1}\eta\right) \times e^{-(\eb,\Delta^{-1} \eta)}.
\label{gen_func_gf}
\end{split}\end{equation}
Taking the second derivative w.r.t.~the source fields and setting $\eb=\eta=0$ yields
\begin{align}
G=\Delta^{-1} + \Delta^{-1} n G_a n \Delta^{-1},
\end{align}
or
\begin{align}
G_a= n^{-1}\Delta G\Delta n^{-1} - n^{-1} \Delta n^{-1}.
\end{align}
This translates into a physical self energy
\begin{align}
\Sigma = g^{-1} - \left[\Delta^{-1} + \Delta^{-1} n G_a n \Delta^{-1}\right]^{-1},
\end{align}
which can be simplified to
\begin{align}
\Sigma = \Sigma_R + \frac{G_R^{-1}n^{-1}\Sigma_a}{G_R n + n^{-1} \Sigma_a},
\label{rel_sig_app}
\end{align}
as reported in \ceq{rel_sig}.

After deriving the relation (\ref{gen_func_gf}) between
the generating functionals of the physical and auxiliary Green functions, we can now establish corresponding relations for 
the effective actions.
Taking the logarithm of the above equation yields
\begin{equation}\begin{split}
\mathcal{W}(\eb,\eta)&=\mathcal{W}_a\left(n^{T}\Delta^{-T}\eb,n\Delta^{-1}\eta\right) - {(\eb,\Delta^{-1} \eta)}
\label{gen_func_conn}
\end{split}\end{equation}
from which we get 
\begin{align}\nonumber
\mathcal{V}(\nub,\nu)&=\left[ \mathcal{W}(\eb,\eta)+(\eb,g\eta)\right]_{\substack{\hspace{-0.15cm}\eta=g^{-1}\nu\\ \eb=g^{-T}\nub}}
=\left[\mathcal{W}_a\left(n^{T}\Delta^{-T}\eb,n\Delta^{-1}\eta\right) - {(\eb,\Delta^{-1} \eta)}+(\eb,g\eta)\right]_{\substack{\hspace{-0.15cm}\eta=g^{-1}\nu\\ \eb=g^{-T}\nub}} \\ \nonumber
&=\left[\mathcal{W}_a\left(\eb,\eta\right)  +(\eb,g_a\eta) -(\eb,g_a\eta) \right]_{\substack{\hspace{-0.4cm}\eta=n\Delta^{-1}g^{-1}\nu\\ \eb=n^{T}\Delta^{-T}g^{-T}\nub}} + \left(\nub,[g^{-1}-g^{-1}\Delta^{-1}g^{-1} ]\nu \right)\\ \nonumber
&=\left[\mathcal{W}_a\left(\eb,\eta\right)  +(\eb,g_a\eta)\right]_{\substack{\hspace{-0.4cm}\eta=g^{-1}_a g_a n\Delta^{-1}g^{-1}\nu\\ \eb=g^{-T}_a g_a^{T}n^{T}\Delta^{-T}g^{-T}\nub}} - (n^{T}\Delta^{-T}g^{-T}\nub,g_an\Delta^{-1}g^{-1}\nu)+ \left(\nub,[g^{-1}-g^{-1}\Delta^{-1}g^{-1}]\nu \right)\\
&=\mathcal{V}_a(g_a^{T}n^{T}\Delta^{-T}g^{-T}\nub,g_an\Delta^{-1}g^{-1}\nu) - (n^{T}\Delta^{-T}g^{-T}\nub,g_an\Delta^{-1}g^{-1}\nu) + \left(\nub,[g^{-1}-g^{-1}\Delta^{-1}g^{-1}]\nu \right).
\end{align}
Some algebra yields
\begin{align}\nonumber
\nu^*&:=g_an\Delta^{-1}g^{-1}\nu = -n^{-1}G_R^{-1}\frac{1}{1-g\Sigma_R} \nu \\
\nub^*&:=g_a^{T}n^{T}\Delta^{-T}g^{-T}\nub = -n^{-T}G_R^{-T}\frac{1}{1-g^{T}\Sigma_R^T} \nub,
\end{align}
and finally\cite{Katanin2013}
\begin{equation}\begin{split}
\mathcal{V}(\nub,\nu)=\mathcal{V}_a(\nub^*,\nu^*) - \left(\nub, \frac{\Sigma_R}{\mathds{1}-g\Sigma_R} \nu\right).      
\end{split}\end{equation}
Aside from the relation of the physical and auxiliary self-energy, a corresponding relation for the respective 1PI two-particle vertices is easily derived by making use of relation (\ref{gen_func_conn}). 
Taking the fourth derivative w.r.t.~the source fields we find
\begin{equation}\begin{split}
G^{(4),c}&= G^{(4),c}_a \left[\Delta^{-1}n ,\Delta^{-1}n,n\Delta^{-1},n\Delta^{-1}\right].
\end{split}\end{equation}
Amputating the full one-particle Green functions yields
\begin{equation}\begin{split}
\gamma_2 &=  \gamma_{2,a} \left[ G^{-1}\Delta^{-1}nG_a,G^{-1}\Delta^{-1}nG_a, G_an\Delta^{-1}G^{-1} , G_an\Delta^{-1}G^{-1}\right],
\end{split}\end{equation}
which simplifies to
\begin{equation}\begin{split}
\gamma_2 &=  \gamma_{2,a}[\zeta,\zeta,\bar\zeta,\bar\zeta].
\label{relation_gammas}
\end{split}\end{equation}

\subsection{Connection between fRG and the flow in the auxiliary space}
To understand how the flow for the auxiliary fields is related to the conventional fRG flow, we will in the following derive the relations between the flow equations. For this we assume that the scale dependence in $Q_a^\Lambda$ is governed by a physical cutoff $Q^\Lambda$ only. The self-energy relation (\ref{rel_sig_app}) holds also in the scale-dependent case, and we find
\begin{align}\nonumber
\Sd^\Lambda &= G_R^{-1}n^{-1}\Sd_a^\Lambda  \left(G_R n + n^{-1} \Sigma_a^\Lambda\right)^{-1} - G_R^{-1}n^{-1}\Sigma_a^\Lambda \left(G_R n + n^{-1} \Sigma_a^\Lambda\right)^{-1} n^{-1}\Sd_a^\Lambda \left(G_R n + n^{-1} \Sigma_a^\Lambda\right)^{-1}\\ \nonumber
&= G_R^{-1}n^{-1} \left[1 - \Sigma_a^\Lambda \left(G_R n + n^{-1} \Sigma_a^\Lambda\right)^{-1} n^{-1} \right] \Sd_a^\Lambda \left(G_R n + n^{-1} \Sigma_a^\Lambda\right)^{-1} \\  \nonumber
&= \left[G_R^{-1}n^{-1} - \left(\Sigma^\Lambda-\Sigma_R\right) n^{-1} \right] \Sd_a^\Lambda \left(G_R n + n^{-1} \Sigma_a^\Lambda\right)^{-1}
\\ \nonumber
&= \left[G_R^{-1}n^{-1} - \left(\Sigma^\Lambda-\Sigma_R\right) n^{-1} \right] \Sd_a^\Lambda \left[G_R n + G_R \left[\left(\Sigma^\Lambda - \Sigma_R\right)^{-1} - G_R\right]^{-1} G_R n\right]^{-1}\\ 
&= \left(Q_R - \Sigma^\Lambda\right) n^{-1} \Sd_a^\Lambda n^{-1} \left(Q_R - \Sigma^\Lambda\right)
=\zeta^\Lambda \Sd_a^\Lambda \bar{\zeta}^\Lambda,
\end{align}
with
\begin{equation}\bar{G}^\Lambda= \left[Q_R-\Sigma^\Lambda \right]^{-1}, \end{equation}
where we have already included a possible scale-dependence in the factors $\zeta^\Lambda=\left(n \bar{G}^\Lambda\right)^{-1}$ and $\bar\zeta^\Lambda=\left(\bar{G}^\Lambda n\right)^{-1}$.
It can also be shown easily that
\begin{equation}
G^\Lambda_a= n^{-1} (Q_R - \Sigma^\Lambda) \left(\frac{1}{Q-\Sigma^\Lambda} - \frac{1}{Q_R-\Sigma^\Lambda}\right) (Q_R - \Sigma^\Lambda)  n^{-1}= \zeta^\Lambda \left(G^\Lambda - \bar{G}^\Lambda \right) \bar{\zeta}^\Lambda.
\label{prop_diff_a}
\end{equation}
Let us now proceed with the single-scale propagator $S_a^\Lambda = \partial_\Lambda G_a^\Lambda |_{\Sigma_a\hspace{0.05cm}\mathrm{fixed}}$. Note that keeping the auxiliary self energy fixed in the derivative is equivalent to keeping the physical self energy fixed. Therefore by taking the derivative of \ceq{prop_diff_a} we find
\begin{align}
S_a^\Lambda = \zeta^\Lambda \left( \partial_\Lambda G^\Lambda |_{\Sigma^\Lambda \phantom{,} \mathrm{fixed}}\right) \bar{\zeta}^\Lambda =\zeta^\Lambda S^\Lambda \bar{\zeta}^\Lambda,
\label{ssc_prop_a}
\end{align}
as $\bar{G}$ and thus $\zeta^\Lambda$ only depend on $\Lambda$ through the self energy.
Looking at the flow in the physical system we find that any vertex in the diagrams effectively gets $\left(n \bar{G}^\Lambda\right)^{-1}$ appended at each connection point and internal propagators are given by $G^\Lambda - \bar{G}^\Lambda$.
Let us now consider the fRG truncation at second order, thus including the vertex flow. 
To achieve this we solve the scale-dependent version of \ceq{relation_gammas} for $\gamma_{2,a}^\Lambda$
\begin{equation}
\gamma_{2,a}^\Lambda =  \gamma_{2}^\Lambda\left[\left( \zeta^\Lambda \right)^{-1},\left( \zeta^\Lambda \right)^{-1},\left(\bar{\zeta}^\Lambda \right)^{-1},\left(\bar{\zeta}^\Lambda \right)^{-1}\right].
\end{equation}
When constructing the diagrams contributing to the auxiliary flow, the $\left( \zeta^\Lambda \right)^{-1}$ factors exactly cancel the outer expressions in $G_a^\Lambda$ of (\ref{prop_diff_a}) and $S_a^\Lambda$ of (\ref{ssc_prop_a}) that connect to a vertex. The diagrams for the flow can thus be translated according
\begin{equation}\begin{split}
\gamma_{2,a}^\Lambda &\rightarrow \gamma_2^\Lambda \\
S^\Lambda_a &\rightarrow S^\Lambda\\
G^\Lambda_a &\rightarrow G^\Lambda - \bar{G}^\Lambda
\end{split}\end{equation}
and the orders in $\gamma_{2,a}$ translate to the corresponding orders in $\gamma_{2}$. 
Let us now come back to the task of deriving a connection between the flow equations in physical and auxiliary space. 
To this end, we start by looking at the flow equation for the two particle vertex. Since we also want to understand the effect of the truncation we first consider the exact flow equations (i.e., without truncation) that involve the 1PI three-particle vertex. This reads, respectively for the physical and auxiliary fermions:
\begin{align} 
\label{gam_flow_notrunc}
\dot{\gamma}_2^\Lambda &= \gamma_2^\Lambda \circ \left( S^\Lambda \circ G^\Lambda + G^\Lambda \circ S^\Lambda \right) \circ \gamma_2^\Lambda+\gamma_3^\Lambda\circ S^\Lambda, \\ 
\dot{\gamma}_{2,a}^\Lambda &= \gamma_{2,a}^\Lambda \circ \left( S_a^\Lambda \circ G_a^\Lambda + G_a^\Lambda \circ S_a^\Lambda \right) \circ \gamma_{2,a}^\Lambda+\gamma_{3,a}^\Lambda\circ S^\Lambda_a. \label{gam_flow_aux_not}
\end{align}
Differentiating \ceq{relation_gammas} with respect to $\Lambda$, substituting \ceq{gam_flow_aux_not}, and using Eqs. (\ref{ssc_prop_a}), (\ref{prop_diff_a}) and (\ref{relation_gammas}) we obtain: 
\begin{align} \nonumber
\dot{\gamma}_2^\Lambda =& \phantom{,}\dot{\gamma}_{2,a}^\Lambda\left[\zeta^\Lambda,\zeta^\Lambda,\bar\zeta^\Lambda,\bar\zeta^\Lambda\right] 
- \gamma_{2,a}^\Lambda\left[\dot{\Sigma}^\Lambda n^{-1},\zeta^\Lambda,\bar\zeta^\Lambda,\bar\zeta^\Lambda\right]
- \gamma_{2,a}^\Lambda \left[\zeta^\Lambda,\dot{\Sigma}^\Lambda n^{-1} ,\bar\zeta^\Lambda,\bar\zeta^\Lambda\right] \nonumber \\ \nonumber
&- \gamma_{2,a}^\Lambda \left[\zeta^\Lambda,\zeta^\Lambda,n^{-1}\dot{\Sigma}^\Lambda,\bar\zeta^\Lambda\right]- \gamma_{2,a}^\Lambda \left[\zeta^\Lambda,\zeta^\Lambda,\bar\zeta^\Lambda,n^{-1}\dot{\Sigma}^\Lambda \right] \\ \nonumber
=& \phantom{,}\gamma_2^\Lambda \circ \left( S^\Lambda \circ \left(G^\Lambda-\bar{G}^\Lambda\right) +\left(G^\Lambda-\bar{G}^\Lambda\right) \circ S^\Lambda \right) \circ \gamma_2^\Lambda +
 \gamma_{3,a}^\Lambda \left[\zeta^\Lambda,\zeta^\Lambda,\zeta^\Lambda,\bar\zeta^\Lambda,\bar\zeta^\Lambda,\bar\zeta^\Lambda \right] \circ S^\Lambda \nonumber \\ 
&-  \gamma_{2}^\Lambda \left[ \dot{\Sigma}^\Lambda \bar{G}^\Lambda,\mathds {1},\mathds {1},\mathds {1}\right]
-  \gamma_{2}^\Lambda \left[ \mathds {1},\dot{\Sigma}^\Lambda \bar{G}^\Lambda,\mathds {1},\mathds {1}\right]
-  \gamma_{2}^\Lambda \left[ \mathds {1},\mathds {1}, \bar{G}^\Lambda\dot{\Sigma}^\Lambda,\mathds {1}\right]
-  \gamma_{2}^\Lambda \left[ \mathds {1},\mathds {1},\mathds {1}, \bar{G}^\Lambda\dot{\Sigma}^\Lambda\right]\label{all_terms}
   \end{align}
Neglecting the term proportional to $\gamma_{3,a}^\Lambda$, consistently with a one-loop approximation in auxiliary space, one directly obtains \ceq{new_gam_flow}. 
To understand the connection between the three-particle physical and auxiliary vertexes, instead, let us keep all the terms and compare \ceq{all_terms} with \ceq{gam_flow_notrunc}. 
A moment of inspection shows that:  
\begin{align}
\gamma_3^\Lambda \circ S^\Lambda  =&\phantom{,} \gamma_{3,a}^\Lambda \left[\zeta^\Lambda,\zeta^\Lambda,\zeta^\Lambda,\bar\zeta^\Lambda,\bar\zeta^\Lambda,\bar\zeta^\Lambda \right] \circ S^\Lambda \nonumber  -
 \gamma_2^\Lambda \circ \left( S^\Lambda \circ  \bar{G}^\Lambda +\bar{G}^\Lambda \circ S^\Lambda \right) \circ \gamma_2^\Lambda \nonumber \\  
&-  \gamma_{2}^\Lambda \left[ \dot{\Sigma}^\Lambda \bar{G}^\Lambda,\mathds {1},\mathds {1},\mathds {1}\right]
-  \gamma_{2}^\Lambda \left[ \mathds {1},\dot{\Sigma}^\Lambda \bar{G}^\Lambda,\mathds {1},\mathds {1}\right]
-  \gamma_{2}^\Lambda \left[ \mathds {1},\mathds {1}, \bar{G}^\Lambda\dot{\Sigma}^\Lambda,\mathds {1}\right]
-  \gamma_{2}^\Lambda \left[ \mathds {1},\mathds {1},\mathds {1}, \bar{G}^\Lambda\dot{\Sigma}^\Lambda\right]  \end{align}
Using the flow equation for the self energy in physical space, \ceq{sig_flow}, and \ceq{ssc_prop_a} yields:
\begin{align}\gamma_3^\Lambda \circ S^\Lambda  =&\phantom{,} \gamma_{3,a}^\Lambda \left[\zeta^\Lambda,\zeta^\Lambda,\zeta^\Lambda,\bar\zeta^\Lambda,\bar\zeta^\Lambda,\bar\zeta^\Lambda \right] \circ S^\Lambda \nonumber  -
 \gamma_2^\Lambda \circ \left( S^\Lambda \circ  \bar{G}^\Lambda +\bar{G}^\Lambda \circ S^\Lambda \right) \circ \gamma_2^\Lambda \nonumber \\ &-  
   \gamma_{2}^\Lambda \left[ S^\Lambda\circ\gamma_2^\Lambda\circ\bar{G}^\Lambda,\mathds {1},\mathds {1},\mathds {1}\right]
-  \gamma_{2}^\Lambda \left[ \mathds {1},S^\Lambda\circ\gamma_2^\Lambda\circ\bar{G}^\Lambda,\mathds {1},\mathds {1}\right] \nonumber \\ 
&-  \gamma_{2}^\Lambda \left[ \mathds {1},\mathds {1}, \bar{G}^\Lambda\circ\gamma^\Lambda_2\circ S^\Lambda,\mathds {1}\right]
-  \gamma_{2}^\Lambda \left[ \mathds {1},\mathds {1},\mathds {1}, \bar{G}^\Lambda\circ\gamma^\Lambda_2\circ S^\Lambda\right]
\label{gam3Der} \end{align}
which can be depicted diagrammatically as shown in \cfg{splitS}.
\begin{figure}
\includegraphics{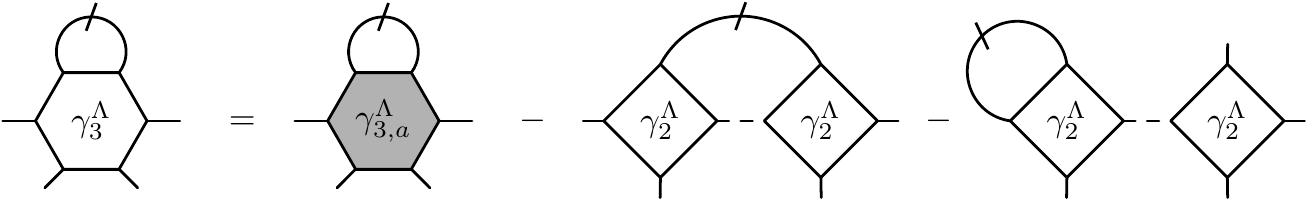}
\caption{Diagrammatic structure of \ceq{gam3Der}. The factors $\zeta^\Lambda$ appended at each leg of $\gamma_{3,a}^\Lambda$ are implicitly included in the diagram.}
\label{splitS}
\end{figure}
The last six terms on the right hand side of this equation can be seen diagrammatically as the possible distinct ways of connecting two out of the six external points of the quantity $\gamma_2^\Lambda \circ \bar G^\Lambda \circ \gamma_2^\Lambda$, and therefore acquire the same diagrammatic structure of the first two terms. Hence by performing a functional derivative with respect to $S^\Lambda$ we can finally write a relation between the three-particle vertex: 
\begin{equation}
\gamma_3^\Lambda=\gamma^\Lambda_{3,a} \left[\zeta^\Lambda,\zeta^\Lambda,\zeta^\Lambda,\bar\zeta^\Lambda,\bar\zeta^\Lambda,\bar\zeta^\Lambda \right]  -\gamma^\Lambda_2\circ \bar G^\Lambda \circ  \gamma^\Lambda_2, 
\end{equation}
as depicted diagrammatically  in \cfg{vsplit}. 
Let us also note explicitly that this equation is consistent with the fact that in the beginning of the flow the auxiliary 1PI three-particle vertex is equal to the amputated connected three particle Green's function of the reference system. 
 
 As for the flow of the self energy one can see from \ceq{sig_sig}, that no further diagrams appear  by performing the flow in the auxiliary space, i.e., the self energy flow equation \ceq{sig_flow} remains the same, and the only differences arise indirectly due to the change of the vertex during the flow discussed above.

\end{widetext}

\bibliography{refs}

\vfill\eject

\end{document}